# Anomaly-based Intrusion Detection System Using Fuzzy Logic

Mohammad Almseidin
*Information Technology Department*
*University of Miskolc*
Miskolc, Hungary
alsaudi@iit.uni-miskolc.hu

Jamil Al-Sawwa
*Computer Science Department*
*Tafila Technical University*
Tafila, Jordan
Jalsawwa@ttu.edu.jo

Mouhammd Alkasassbeh
*Computer Science Department*
*Princess Sumaya University for Technology*
Amman, Jordan
M.alkasassbeh@psut.edu.jo

*Abstract*—Recently, the Distributed Denial of Service (DDOS) attacks has been used for different aspects to denial the number of services for the end-users. Therefore, there is an urgent need to design an effective detection method against this type of attack. A fuzzy inference system offers the results in a more readable and understandable form. This paper introduces an anomaly-based Intrusion Detection (IDS) system using fuzzy logic. The fuzzy logic inference system implemented as a detection method for Distributed Denial of Service (DDOS) attacks. The suggested method was applied to an open-source DDOS dataset. Experimental results show that the anomaly-based Intrusion Detection system using fuzzy logic obtained the best result by utilizing the InfoGain features selection method besides the fuzzy inference system, the results were 91.1% for the true-positive rate and 0.006% for the false-positive rate.

## I. INTRODUCTION

It seems impossible to say that intrusions are not increased and developed continuously. Every day, there is a large amount of financial loss, privacy violation, and information transfers in an illegal way, as a result of succeeding intrusions implementation. There are different types of intrusions threatening networks, computer information, and resources. Each type of intrusions aims to implement specific tasks. As well as, under the umbrella of each intrusions type there are several intrusions with different techniques, share a similar purpose.

As an example, Trojan attacks implemented as a daemon program installed in the background without legal permission. From another perspective, remote to local intrusions concentrates on giving root grant permissions for intruders. The previous types of intrusions could be implemented as prerequisite steps to implement the Denial Of Service (DOS) intrusions. This type of intrusion heading consume various resources in order to close down several services for legal users. The distribution of DOS attacks represents the 60% of a total number of attacks around the world [1]–[4].

An intrusion Detection System (IDS) is one of the effective solutions to detect and prevents intrusions occurrence. Hence, the urgency to build an effective intrusion detection system is publicly needed. In another dimension, there are different challenges to implement a sufficient IDS. One of these challenges is the crisp mechanism of detection techniques. In other words, the typical detection mechanisms of IDS had a boundary problem [5]–[8]. The boundary problem appears in IDS because there are different type of intrusions cannot be expressed and detected in crisply fashion [9]–[11]. In general, the fuzzy system offers several advantages to handle crisp boundary problems. It also presents the detection degree level of intrusions which could be more readable for the security engineer. In this work, we break down the implementation of the detection mechanism for DDOS using fuzzy logic into two challenging steps. Step 1: To identify how the fuzzy system can be used for security purposes and how it can handle the intrusion boundary problem. Step 2: To implement the classical fuzzy inference system as a detection mechanism for the recent types of DDOS attacks.

The rest of the paper is organized as follows: Section II illustrates recent works related to the detection methods of the distributed denial of service attacks. Section III introduces a brief overview of the distributed denial of service attack types. Section V presents the proposed detection approach in detail followed by the results and discussions in subsection C. Finally, section VI concludes the paper.

## II. RELATED WORKS

This section presents the related works relevant to using the fuzzy system for intrusion detection. It also provides a brief overview of different methods and approaches that are used for intrusion detection.

In [12], authors implement an architecture to detect DDOS attack using Fuzzy Reasoning Spiking Neural-P (FRSN-P). It is a type of membrane computing system. The neurons within this system communicate based on electrical spikes (impulses). The Knowledge Discovery Databases (KDD-99) dataset imported into the proposed system. KDD-99 dataset was prepared by the University of California [13]. The constraint was on a synchronization flood. The authors extracted the synchronization flood attack from the KDD-99 dataset. After extracting the required records of the desired attack. The fuzzy reasoning spiking neural-P implemented and evaluated. According to the registered experiments, the proposed system reached effectively the following values 0.02% and 0.25% for false-negative rate and false positive rate respectively.

On the other hand, authors in [5], proposed a network intrusion detection system based on automatic fuzzy rule base generation. The single length of frequent item approach was implemented as prepossess step to generate the required fuzzy rule base automatically. KDD-99 dataset was used to evaluate the proposed system. The implemented experiments recorded the success of detection of different types of DDOS attacks within the imported dataset, simultaneously, the proposed system recorded 90% as overall accuracy rate.

Further, the effort of [14], focuses on detecting and prevent the Neptune attack. It is a type of TCP flood attack and belongs to DOS attacks. The enhanced release of the KDD99 dataset (NSL-KDD) was imported to test and evaluate the proposed fuzzy system. According to the implementation of the proposed fuzzy system, a features ranking algorithm was implemented to select the relevant features for the detection approach. Further, the proposed fuzzy system was evaluated with a decision tree algorithm using the same NSLKDD dataset. As a result of implementing and evaluating the proposed fuzzy system with a decision tree algorithm. The proposed fuzzy system succeeded to obtain 0.93% as the overall average accuracy rate for detecting Neptune attack. Comparing with the decision tree algorithm, the proposed fuzzy system had the highest average accuracy rate.

There are different methods that have been used for IDS to implement the feature selection algorithms. There are a large number of features can be collected using several network tools. However, not all of them relevant to the detection mechanism. Therefore, reducing the number of irrelevant features helps to improve the efficiency of the proposed system. As it relates to the features selection algorithms. In [5], the authors proposed a network intrusion detection system based on the fuzzy system. The feature selection algorithm was implemented to reduce a large number of features (41) that belong to KDD-99.

The authors divided the imported dataset into two parts, the first part for the training phase and the second part for the testing phase. The fuzzy rules generated based on knowledge experts. As a result of the proposed fuzzy system experiments, the average accuracy rate recorded as 0.95% with an acceptable false-positive rate.

From another perspective, different algorithms have been used in parallel with a fuzzy system to detect and prevent intrusions. In the efforts of [15], a hybrid approach of genetic algorithm and the fuzzy system was implemented. The main objective of the proposed system to detect DOS attacks. The main purpose of implementing a genetic algorithm to generate the fuzzy rules automatically. Therefore, the genetic algorithm presented as preprocess step of the proposed system. The testbed environment was implemented using the KDD-99 dataset. The proposed system recorded 0.94% as the average detection rate.

As well, there are other efforts suggesting neural network algorithms as a hybrid approach with a fuzzy system. In [16], the authors incorporate the neural network algorithm and fuzzy system to enhance the detecting rate of the intrusion detection system. The fuzzy rule base generated using classification taskparallel with the knowledge base. This incorporation reached 0.93% as the average of detection. From another dimension, in [17], incorporation between fuzzy system and decision tree algorithm was implemented and evaluated. The main contribution of the implementing decision tree algorithm is to extract the more significant fuzzy rule based on the previous step of feature selection. The proposed system was evaluated and compared with another hybrid intelligent approach. Consequently, The required improvement of detection rate was obtained. As well as; a study in [18], authors focused on enhancing the detection rate of an intrusion detection system based on combining the fuzzy system and particle swarm optimization (PSO) method. The PSO method was implemented to generate a fuzzy rule base in order to detect DDOS attacks. As a result of, the generated fuzzy rules, the proposed fuzzy system was build and the result reached 0.93% as an average detection rate.

The past works provided convincing contributions and at the same time support, the idea of implementing the fuzzy system produced a rightful approach for intrusions detection. From another perspective, the previous methods still have common gaps, such as it did not includes the recent types of DDOS attacks, for instance, HTTP attacks because they are using either the KDD-99 dataset or NSL KDD-99 dataset. In this work, the recent types of DDOS attacks such as HTTP flood and SQL injection attack were included. Moreover, the fuzzy logic implemented as a detection mechanism for DDOS attacks.

## III. DISTRIBUTED DENIAL OF SERVICE ATTACKS

DDOS attack treated as one of the most harmful types of attack. Nowadays, DDOS attacks are considered a continuous challenge for both users and organizations. The first serious appearance of DDOS was in 2000 and Yahoo was the first victim of DDOS [19], [20]. The main purpose of DDOS attacks is to consume different types of resources such as network bandwidth, CPU, memory utilization etc. Any consumption of these resources will increase the overloading and as a result, different services would be unavailable for legal users.

There different types of (DDOS) attacks such as the Smurf attack where an intruder sends large numbers of Internet Control Message Protocol (ICMP) echo packets to the intended victim. In most situations the intermediary(slave machine) does not filter ICMP messages, therefore, many clients on the network who receive this ICMP echo request sent ICMP replay back. Another type of DDOS attack is a User Datagram Protocol (UDP) flood attack. It is one of the most common types of a DDOS attack, where the intruders send a large number of UDP traffics to the victim within a specific period of time. Meanwhile, an HTTP-flood attack considered a difficult one to detect. According to the HTTP-flood attack, the intruder sends completely normal posted messages at a very slow rate in a systematic way, this type of DDOS is difficult to detect because its behavior seems like normal behavior. Another type of modern DDOS attack is a Simple Query Language (SQL) injection Distributed Denial Of Service (SIDDOS). According to this type of DDOS, the intruder inserts a malicious SQL statement as a string in the browser side, then it forwarded to the victim as an executed statement [21].

## IV. DISTRIBUTED DENIAL OF SERVICE DATA-SET

Recently, there are several open-source datasets that include different types of intrusions, such as knowledge discovery in databases and ADFA Intrusion detection datasets. These datasets provide a convenient environment for research purposes [13].

With regards to the previous intrusions dataset. The recent DDOS dataset of Alkasassbeh *et al.* used as a testbed environment.

*A. Dataset Prepossessing and Features Selection*

The first step of preprocessing was extracting 10% of the total number of connections of records. The extracted DDOS dataset categorized into two classes either normal or intrusions with respect to the previous class label. The typical intrusion detection systems detect and classify the packet based on predefined rules such as SNORT [22]. Hence, the predefined rules which are responsible for detecting intrusion or normal packet could be figured based on the given attributes of the DDOS dataset. There are a large number of attributes that could be recorded during the implementation of intrusions datasets. These attributes recorded using any network monitoring tools. From another perspective, generally, most of them are not highly ranked attributes. which is means, these attributes are not related to the detection mechanisms. Attribute selection presented as a very important step because if there are irrelevant attributes and it did not satisfy the aim of the proposed system. As a consequence, Information Gain (IG) which is the common method of attribute selection implemented [23].

## V. FUZZY INFERENCE SYSTEM FOR IDS (FIS-IDS)

Regarding the design and implementation of FIS-IDS, the following prerequisites should be done. The fuzzy sets of input/output. Meanwhile, the membership functions for every input and output of FIS-IDS. Membership functions recognize the degree of participation (degree of membership) for each input/output. They presented graphically to define the expected output response. Furthermore, the degree of membership for each input used in the fuzzy rule base in order to define their impact on the output fuzzy sets. In this work, the input of FISIDS was the extracted DDOS dataset. Regarding determining the range values of membership functions, the maximum, minimum, and average values for each continuous attribute were computed. The membership functions divided into two and three linguistic terms as Table I presented.

The trapezoid membership functions are applied for the inputs of FIS-IDS. The output response of FIS-IDS presents the percentage degree of DDOS intrusion availability. It is divided into the following linguistic terms (False, Low Attack, Medium Attack, and High Attack). Meanwhile, it is distributed from [0 1]. The crisp value of the output response obtained based on the centroid method.

TABLE I
LINGUISTICS TRANSFORMATION

| Attributes Name | Membership Functions |
|---|---|
| PKT SIZE | Small,Medium,Large |
| PKT RATE | Low,High |
| BYTE RATE | Low,High |
| PKT AVG SIZE | Small,Medium,Large |

*A. Estimation of Fuzzification and Membership Functions*

In this work, the test-bed DDOS dataset was divided into training cases and test cases. The training data consisted of 10000 instances of records with 5000 normal cases and 5000 intrusion cases. The test data consisted of 10000 records with 5000 normal cases and 5000 intrusion cases. In order to increase the efficiency of IDS, it is important to identify the observable features that are relevant to detect intrusions from the network traffic data [12]. In spite of, These are some of the relevant features: Pkt Size, Byte Rate, Pkt Avg Size, Pkt Delay, Utilization and Pkt Rate. The most relevant features according to the IG algorithm were used. These features as found in the test-bed DDOS dataset are:

- Pkt Size: It is the feature number (7) as found in the test-bed DDOS dataset. It represents the total packet size in byte the send from the source to the destination.
- Pkt Rate: it is the feature number (19) and illustrates the total packet rate between source and destination.
- Byte Rate: It is the feature number (20) and indicates the total byte rate between source and destination.
- Pkt Avg Size: It is the feature number (21) and represents the average packet size between source and destination.

The anomaly-based and misuse-based techniques of IDS detect attacks based on a predefined rule base( i.e. rules for normal and intrusions). For these, sorting the normal and intrusions records of the dataset is required [24]. The following algorithm illustrates the preprocess step of sorting and features extraction of the test-bed dataset. Therefore, the outputs of the sorting and feature extraction algorithm are two pools of normal and intrusions records besides their ranges (maximum, minimum, and average) values in case of normal, intrusions, and mixed (normal and intrusion). The aim behind stores and investigates the maximum, minimum, and average pools that, the average value of the identified features decreased in the presence of intrusions while it increased for some other identified features [14]. These two pools consisted only of the relevant observable features that are suitable for IDS ($7^{th}$, $19^{th}$, $20^{th}$ and $21^{st}$) features, where all other values removed. The parameters of fuzzy sets were constructed and optimized using the method of [24], the As a result of Algorithm (2), these fuzzy sets were represented by a trapezoidal membership function. ($7^{th}$ and $21^{st}$) features had three trapezoidal membership functions to represent their linguistic terms (small, medium, and large). ($19^{th}$ and $20^{th}$) features had two trapezoidal membership functions to represent their linguistic terms (low and high). following algorithm shows the process of estimation fuzzy sets parameters for every selected relevant feature based on the training dataset.

---
**Algorithm 1** Sorting and Feature Extraction Algorithm
---
Input : The imported test-bed dataset
Output: Two pools of test-bed dataset (normal and intrusions), the maximum, minimum and average values of normal and intrusions pools
1: while Termination Condition Not Met do
2:    Classify whole test-bed dataset into "normal" and "attack" class
3:    Extract the suitable features for IDS based o IG algorithm
4:    Remove all irrelevant features
5:    Store the normal pool
6:    Store the intrusions pool

7: Find the maximum and minimum values for every continues feature of normal and intrusions pools
8:     Store the range values of normal pool
9:     Store the range values of intrusions pool
10: end while

---

Algorithm 2 Fuzzy Sets Construction and Optimization

Input : Normal pool and intrusions pool
Output: Fuzzy sets of FIS-IDS

1: $\gamma$ presents the largest value within selected feature
2: $\beta_1$ and $\beta_2$ present the first and second average of the selected feature
3: $\alpha$ indicates the smallest value of selected the feature
4: Select features from normal pool and intrusions pool
5: Check for missing entry for all records
6: Select records from the normal and intrusions pools
7: Divide each of ($7^{th}$ and $21^{st}$) features into Small, Medium and Large
8: Divide each of ($19^{th}$ and $20^{th}$) features into Low and High
9: Process all selected features
10: Set fuzzy membership value for every feature: $\alpha+\gamma=2\beta$
11: Calculate fuzzy membership value for every continuous feature
12: Store all fuzzy sets
13: Repeat step 9 until all selected columns are covered
14: Repeat step 4 until all records in the normal and intrusions pools are considered

---

### B. Fuzzy Rule Generation

The part of the fuzzy rule generation considered as one of the important parts of the fuzzy system. For the purpose of determining the level of attack, the fuzzy rule base of the FIS-IDS was generated in a simple strategy. In the early stages of the fuzzy rules generation of FIS-IDS, all of the possible permutations of the membership values of the identified input parameters were itemized. The total number of the generated permutations was (36) used to create the fuzzy rule base. These permutations used to represent the antecedent part of the fuzzy rule base. In distinction to the training data, the noticeable changes of the behavior of the average value for each identified feature were observed and investigated. These noticeable changes observed in cases of normal, intrusion, and mixed (normal and intrusion). The consequent part of the fuzzy rule was based on the behavior noticeable changes of the average values, where the average value of the identified features decreased in case of occurrence the intrusions while it increased for some other identified features. Additionally, the generated permutations were divided into three clusters, which presented in

[1] HINT: S = SMALL, M = MEDIUM, L = LARGE, LO = LOW, H = HIGH, FA = FALSE ATTACK, LA = LOW ATTACK, MA = MEDIUM ATTACK, HA = HIGH ATTACK.

Tables II, III, and IV respectively. Every cluster consisted of 12 permutations represent the antecedent part of fuzzy rule, where the consequent part was from the observed behaviour of the average values of identified features [1].

TABLE II
THE CLUSTER.1 OF PERMUTATIONS

|     | Antecedent Part | | | | Consequent Part |
| --- | --- | --- | --- | --- | --- |
| No. | Pkt Size | Byte Rate | Pkt Rate | Pkt Avg | Attack Possibility |
| 1 | S | LO | LO | S | FA |
| 2 | S | LO | LO | M | FA |
| 3 | S | LO | LO | L | LA |
| 4 | S | H | H | S | LA |
| 5 | M | LO | LO | S | FA |
| 6 | M | LO | LO | M | FA |
| 7 | M | LO | LO | L | LA |
| 8 | M | H | H | S | MA |
| 9 | L | LO | LO | S | FA |
| 10 | L | LO | LO | M | LA |
| 11 | L | LO | LO | L | MA |
| 12 | L | H | H | S | MA |

TABLE III
THE CLUSTER.2 OF PERMUTATIONS

|     | Antecedent Part | | | | Consequent Part |
| --- | --- | --- | --- | --- | --- |
| No. | Pkt Size | Byte Rate | Pkt Rate | Pkt Avg | Attack Possibility |
| 13 | S | H | H | M | MA |
| 14 | S | H | H | L | HA |
| 15 | S | LO | H | S | FA |
| 16 | S | LO | H | M | LA |
| 17 | M | H | H | M | MA |
| 18 | M | H | H | L | HA |
| 19 | M | LO | H | S | LA |
| 20 | M | LO | H | M | MA |
| 21 | L | H | H | M | MA |
| 22 | L | H | H | L | HA |
| 23 | L | LO | H | S | LA |
| 24 | L | LO | H | M | MA |

The fuzzy rules were designed based on expert knowledge of detecting DDOS attacks, obtained range values, and deeply studies and investigate the relationship between the most relevant features and detecting DDOS intrusions, Table V shows the extracted fuzzy rules.

### C. Experiments and Results of FIS-IDS

Thereupon, all experiments of FIS-IDS were conducted using Matlab, Matlab Fuzzy Toolbox, and the Mamdani inference system. The Mamdani inference system was implemented for the reasoning part and alert generation [25]. The crisp value of the output response obtained using the centroid method. Meanwhile, it presents the percentage degree of attack availability. The imported DDOS dataset was divided into three parts as follows:

- The first part of the imported dataset was used to design the FIS-IDS, optimizing membership functions and validity check.
- The second part of the DDOS dataset was used for testing and evaluating the performance metrics of detecting intrusions records.
- The third part of the DDOS dataset used for testing and evaluating the performance metrics of detecting normal records.

TABLE IV
THE CLUSTER.3 OF PERMUTATIONS

| | Antecedent Part | | | | Consequent Part |
|---|---|---|---|---|---|
| No. | Pkt Size | Byte Rate | Pkt Rate | Pkt Avg | Attack Possibility |
| 25 | S | LO | H | L | MA |
| 26 | S | H | LO | S | FA |
| 27 | S | H | LO | M | LA |
| 28 | S | H | LO | L | MA |
| 29 | M | LO | H | L | HA |
| 30 | M | H | LO | S | LA |
| 31 | M | H | LO | M | MA |
| 32 | M | H | LO | L | HA |
| 33 | L | LO | H | L | HA |
| 34 | L | H | LO | S | LA |
| 35 | L | H | LO | M | MA |
| 36 | L | H | LO | L | HA |

TABLE V
MEMBERSHIP TERMS OF FIS-IDS FUZZY RULES

| | Antecedent Part | | | | Consequent Part |
|---|---|---|---|---|---|
| Index | Pkt Size | Byte Rate | Pkt Rate | Pkt Avg | Attack Possibility |
| 1 | S | LO | LO | S | FA |
| 4 | S | H | H | S | LA |
| 20 | M | LO | H | M | MA |
| 23 | M | H | LO | M | MA |
| 27 | L | LO | LO | L | MA |
| 24 | L | H | LO | L | HA |
| 33 | L | LO | H | L | HA |
| 32 | L | LO | H | M | MA |
| 14 | M | LO | LO | M | FA |
| 25 | L | LO | LO | S | FA |
| 18 | M | H | H | L | HA |
| 31 | L | LO | H | S | LA |
| 15 | M | LO | LO | L | LA |
| 6 | S | H | H | L | HA |

Additionally, the inputs of the fuzzy inference system are the raw data (Packet size, Packet rate, Byte rate, and Average packet size). We extracted 5000 instances of normal records and 5000 instances of intrusions records. Regarding the normal test scenario, 5000 instances of normal records are used as inputs of the fuzzy inference system. In order to check the validity of outputs results. We have used the fuzzy inference system evaluation script. It provides the capabilities to test a large number of observations. The result of the FIS evaluation script of the normal test matrix was recorded. It presents the output response of the FIS-IDS which indicates the percentage degree of intrusion availabilities. From another perspective, the intrusions test scenario implemented using 5000 instances of intrusions records, where the intrusion test matrix presents the inputs of FIS-IDS. Moreover, the fuzzy inference system evaluation script was executed for the intrusion test matrix. The two outputs matrices of the normal test scenario and intrusion test scenario are recorded and compared with the actual normal records and intrusions records. There are several measurement metrics that could be calculated from the outputs matrices. According to [26] the typical IDS performance metrics are: True Positive (TP), True Negative (TN), False Negative (FN), False Positive (FP), and Detection Rate (DR), thus, these measurement metrics are computed.

TABLE VI
THE TOTAL NUMBER OF TESTED RECORDS OF FIS-IDS

| | Normal | Intrusion | Total |
|---|---|---|---|
| Normal | 4970 | 30 | 5000 |
| Intrusion | 445 | 4555 | 5000 |
| Total | 5415 | 4585 | 10000 |

According to the two output matrices of normal and intrusions test scenarios. There is 10000 instance of records tested and evaluated using classical fuzzy inference system. The total number of tested records for both normal and intrusions records listed in Table VI. Meanwhile, Table VII illustrates the computed confusion matrix of FIS-IDS.

TABLE VII
CONFUSION MATRIX OF FIS-IDS

| Alert Response | Normal Packet Prediction | Intrusion Packet Prediction |
|---|---|---|
| Normal | TPR = 0.911 | FPR = 0.006 |
| Intrusions | FNR = 0.089 | TNR = 0.994 |

## VI. CONCLUSION

This paper introduces an anomaly-based Intrusion Detection system using fuzzy logic. The suggested method was applied to detect the Distributed Denial of Service (DDOS) attacks. The info-gain features selection algorithm was used to select the relevant features from the DDoS-2016 dataset. These relevant features are packet size, packet rate, byte rate, and the average packet size. The evaluation procedure was carried out by extracting 5000 instances of normal records and 5000 instances of intrusions records. Regarding the normal test scenario, 5000 instances of normal records are used as inputs of the fuzzy inference system. In order to check the validity of outputs results. We have used the fuzzy inference system evaluation script. The two outputs matrices of the normal test scenario and intrusion test scenario are recorded and compared with the actual normal records and intrusions records. The experiment results show that the suggested method obtained 91.1% as true positive rate, and 0.006 for the false positive rate. Furthermore, the false-negative rate was 0.089.